\documentclass[preprint,5p,number,sort&compress]{elsarticle}

\usepackage[english]{babel}
\usepackage[utf8]{inputenc}
\usepackage[T1]{fontenc}
\usepackage{amsmath}
\usepackage{graphicx}
\usepackage[colorinlistoftodos]{todonotes}
\usepackage{hyperref}
\usepackage{dcolumn}
\usepackage[exponent-product = \cdot, product-units = power, separate-uncertainty = true]{siunitx}
\usepackage{booktabs}
\usepackage{lineno}

\begin{document}

\begin{frontmatter}

\title{PENTrack---a simulation tool for ultracold neutrons, protons, and electrons in complex electromagnetic fields and geometries}

\author[TUM]{W. Schreyer\corref{cor1}}
\ead{w.schreyer@tum.de}
\author[TRIUMF]{T. Kikawa}
\author[TUM]{M. J. Losekamm}
\author[TUM]{S. Paul}
\author[TRIUMF,SFU]{R. Picker}

\address[TUM]{Technical University of Munich, James-Franck-Str. 1, 85748 Garching, Germany}
\address[TRIUMF]{TRIUMF, 4004 Wesbrook Mall, Vancouver, Canada}
\address[SFU]{Simon Fraser University, 8888 University Drive, Burnaby, Canada}
\cortext[cor1]{Corresponding author}

\date{\today}

\begin{abstract}
Modern precision experiments trapping low-energy particles require detailed simulations of particle trajectories and spin precession to determine systematic measurement limitations and apparatus deficiencies. We developed PENTrack, a tool that allows to simulate trajectories of ultracold neutrons and their decay products---protons and electrons---and the precession of their spins in complex geometries and electromagnetic fields. The interaction of ultracold neutrons with matter is implemented with the Fermi-potential formalism and diffuse scattering using Lambert and microroughness models. The results of several benchmark simulations agree with STARucn v1.2, uncovered several flaws in \textsc{Geant4} v10.2.2, and agree with experimental data. Experiment geometry and electromagnetic fields can be imported from commercial computer-aided-design and finite-element software. All simulation parameters are defined in simple text files allowing quick changes. The simulation code is written in C++ and is freely available at \href{https://github.com/wschreyer/PENTrack.git}{\nolinkurl{github.com/wschreyer/PENTrack.git}}.
\end{abstract}


\end{frontmatter}


\section{Motivation}

Precision experiments with particles at low energies require an excellent understanding of particle trajectories. Apparatus effects can influence the measurements and lead to false results.

Measurements of the neutron lifetime are a prime example. Neutron-lifetime experiments storing ultracold neutrons (UCNs) in material bottles recently have suffered from poorly understood apparatus effects \cite{Arzumanov12,Steyerl2012}, e.g. unaccounted losses of UCNs at the bottle walls, and their results often deviate beyond the quoted uncertainties \cite{PDG2016}. They also deviate from results of beam experiments, which determine the neutron-decay rate from cold-neutron beams \cite{Yue2013}.

To improve this situation, next-generation experiments like PENeLOPE \cite{Materne2009176}, UCN\(\tau\) \cite{Salvat2014}, and HOPE \cite{Leung2016} plan to trap UCNs in complicated magnetic-field configurations and plan to detect the decay products---protons and electrons. To study apparatus effects of this type of trap, simulation tools are needed to track neutrons, protons, and electrons in inhomogeneous, time-dependent electromagnetic fields.

The search for an electric dipole moment of the neutron (nEDM) using trapped UCNs may give key constraints to CP-violating mechanisms necessary to explain the matter-antimatter asymmetry of the universe \cite{Morrissey2012}. Several new nEDM experiments are currently under construction. To reach the aspired sensitivity of \SI{e-27}{\elementarycharge\centi\meter}, simulations are needed to study apparatus effects, e.g. geometric phases, and optimize the performance of these experiments \cite{Pendlebury2004,Bodek2011,Katayama2014}.

Existing simulation codes, e.g. STARucn and MCUCN \cite{Clement2014,Bodek2011}, allow to simulate interactions of UCNs with matter and spin precession in weak magnetic fields. However, they cannot calculate UCN trajectories in strong magnetic fields and require a description of experiment geometries based on combinations of basic volumetric shapes, making implementation of complex geometries difficult. \textsc{Geant4} \cite{Agostinelli2003}, based on \cite{Atchison2005}, allows the most comprehensive simulations of UCNs, neutron spins, protons, and electrons in electromagnetic fields.

For the PENeLOPE project, we developed the simulation tool PENTrack. It allows simulations of complete neutron-lifetime and nEDM experiments. The implemented physics processes cover UCN transport, UCN storage in material bottles and magnetic traps, spin precession of neutrons and co-magnetometer atoms, and tracking of protons and electrons in electromagnetic fields. It provides a flexible configuration interface and allows to load complex electromagnetic fields and experiment geometries directly from finite-element (FEM) and computer-aided-design (CAD) software. In this paper, we describe the underlying physics and algorithms, compare our results to experiments and other simulation tools, and provide examples for the optimization of experiments and the estimation of false results due to apparatus effects.

\section{Description}

\subsection{Equation of motion}

Simulations of particles in electromagnetic fields require a numerical integration of their equation of motion. PENTrack performs a error-controlled fifth-order Runge-Kutta integration \cite{odeint} of the relativistic equation of motion,\footnote{\( \dot{\mathbf{x}} = \mathrm{d}\mathbf{x}/\mathrm{d}t \) and \( \ddot{\mathbf{x}} = \mathrm{d}^2\mathbf{x}/\mathrm{d}t^2 \).}
\begin{equation}
\ddot{\mathbf{x}} = \frac{1}{\gamma m} \left( \mathbf{F} - \frac{1}{c^2} \left( \dot{\mathbf{x}} \cdot \mathbf{F} \right) \dot{\mathbf{x}} \right),
\label{eq:eom}
\end{equation}
of a particle with mass \(m\), charge \(q\), magnetic moment \(\mu\), and relativistic Lorentz factor \(\gamma\) in the reference frame of the experiment. The force
\begin{equation}
\mathbf{F} = m\mathbf{g} + q \left(\mathbf{E} + \dot{\mathbf{x}} \times \mathbf{B} \right) + p \mu \boldsymbol{\nabla} \! \left| \mathbf{B} \right|
\end{equation}
includes i) gravitational acceleration \(\left| \mathbf{g} \right| = \SI{9.80665}{\meter\per\second\squared}\) in negative \(z\) direction, ii) Lorentz force of a magnetic field \(\mathbf{B}\) and an electric field \(\mathbf{E}\), and iii) the force of a magnetic gradient \(\boldsymbol{\nabla} \! \left|\mathbf{B}\right|\) on the magnetic moment with a polarization \(p\) of \num{+-1}.


\subsection{Interaction with matter}

Ultracold neutrons strongly interact with matter; interactions of protons and electrons with matter are not implemented yet. The latter particles are considered lost as soon as they hit a surface.

All materials in the simulation are described by a complex optical potential \(U = V - i W\) \cite{Golub}. Its real part,
\begin{equation}
V = \frac{2 \pi \hbar}{m_\mathrm{n}} \sum_i b_i n_i,
\end{equation}
depends on the number densities \(n_i\) and bound coherent scattering lengths \(b_i\) of each nucleus species \(i\). The imaginary part,
\begin{equation}
W = \frac{\hbar}{2} \sum_i n_i \sigma_{l,i} v_\mathrm{n},
\end{equation}
depends on the loss cross sections \(\sigma_{l,i}\) for a given velocity \(v_\mathrm{n}\). This cross section is the sum of absorption and inelastic-scattering cross sections, since inelastic scattering increases the energy of a UCN so far above the storage potential that it can be considered lost. Scattering lengths and absorption cross sections are tabulated in \cite{Sears92}. Inelastic-scattering cross sections, however, are often unknown.

Interaction with the surface of a material can lead to reflection, absorption, or transmission through the surface.

The reflection probability \(R\) for a UCN with kinetic energy \(E\) hitting a surface at an incident angle \(\theta\) depends on the energy component perpendicular to the surface, \(E_\perp = E \cos^2 \theta\):
\begin{equation}
R = \left| \frac{ \sqrt{E_\perp} - \sqrt{E_\perp - V + i W} }{ \sqrt{E_\perp} + \sqrt{E_\perp - V + i W} } \right| ^2.
\end{equation}
If the UCN is not reflected but transmitted through the surface and into the material, the velocity of the UCN undergoes refraction, changing its kinetic energy to \(E' = E - V\). The wave number \(k' = \sqrt{2 m \left( E - V + iW \right)} / \hbar \) becomes complex and leads to an exponential decay of the amplitude \(\exp \left[ -\mathrm{Im}(k') x \right] \). The loss probability after a path length \(d\) in the material is then
\begin{equation}
P_\mathrm{loss} = 1 - \exp \left[ -2 \cdot \mathrm{Im} \left( \frac{\sqrt{2 m \left( E - V + i W \right) }}{\hbar} \right) d \right].
\end{equation}

A UCN impinging on a surface can be scattered specularly or diffusely. PENTrack calculates the scattering distribution of the latter process using either a simple Lambert model or the microroughness model, as introduced in \cite{Steyerl72} and validated in \cite{Atchison10}. To calculate the total microroughness-scattering probability one has to integrate the distribution over all scattering angles. PENTrack uses a fast Gauss-Kronrod integration \cite{alglib}, which only slightly impacts the performance.

\subsection{Spin motion}

Every spin-\(\frac{1}{2}\) particle has a magnetic moment of size \(\mu\) parallel or antiparallel to its spin vector \(\mathbf{S}\). Its motion in the reference frame of the experiment follows the Bargmann-Michel-Telegdi (BMT) equation \cite{Rebilas2011},
\begin{equation}
\dot{\mathbf{S}} = \left( -\frac{2 \mu}{\gamma \hbar} \mathbf{B}' + \boldsymbol{\omega}_T \right) \times \mathbf{S},
\label{eq:BMT}
\end{equation}
which describes a precession around the sum of the magnetic-field vector in the rest frame of the particle,
\begin{equation}
\label{eq:vcrossE}
\mathbf{B}' = \gamma \mathbf{B} + \left( 1 - \gamma \right) \left( \mathbf{B} \cdot \dot{\mathbf{x}} \right) \frac{\dot{\mathbf{x}}}{\dot{\mathbf{x}}^2} - \frac{\gamma}{c^2} \dot{\mathbf{x}} \times \mathbf{E},
\end{equation}
and the Thomas-precession axis,
\begin{equation}
\boldsymbol{\omega}_T = \frac{\gamma^2}{c^2 \left( \gamma + 1 \right)} \ddot{\mathbf{x}} \times \dot{\mathbf{x}}.
\end{equation}
After integrating a step of the particle trajectory, PENTrack separately integrates the BMT equation along this step. The separate integration of trajectory and spin precession allows each numerical integrator to choose the optimal internal step length for each process and improves performance.

Since the BMT equation is only valid in small magnetic fields, the user can define a threshold field only below which the BMT equation is integrated. Once a particle enters a field above this threshold, its spin collapses into one of its fully polarized eigenstates. The probability \(P\) to find the polarization \(p\) being parallel or antiparallel to the magnetic field is given by the projection of the spin onto the magnetic field:
\begin{equation}
P\left(p = \pm 1 \right) = \frac{1}{2} \left( 1 \pm \frac{\mathbf{B}}{\left| \mathbf{B} \right|} \cdot \frac{\mathbf{S}}{\left| \mathbf{S} \right|} \right).
\end{equation}

The sign of the polarization can also be flipped during reflection on surfaces. This can be accounted for by assigning a fixed spin-flip probability to each material.

\subsection{Configuration}

PENTrack allows to load maps of magnetic and electric fields calculated with commercial FEM tools like OPERA \cite{opera} on regularly spaced grids. Both three-dimensional maps of arbitrary fields and two-dimensional maps of rotationally symmetric fields are supported. Field values between grid points are calculated with tri- and bicubic interpolation \cite{Lekien05, alglib}. We also implemented analytic magnetic fields, e.g. nearly homogeneous fields with small gradients and fields of straight conductors. Arbitrary time dependence of the fields can be described by a user-defined function.

Experiment geometries can be imported from virtually all CAD software as StL files \cite{StL89}. StL files approximate surfaces with triangle meshes. The intersection of a particle trajectory with such a surface is detected using the CGAL library \cite{cgal}. Each part of the geometry can be deactivated in user-defined time intervals, making it possible to simulate variable properties of valves and other moving parts.

All simulation parameters---material properties, geometric model files, field maps, particle sources, and particle spectra---are stored in simple text files, allowing quick changes and optimization. Random samples from initial distributions and random processes are generated by a Mersenne Twister random-number generator, provided by the Boost libraries \cite{Boost2016}.

Several variables of each tracked particle can be recorded: its position, velocity, and polarization at beginning and end of the simulation, at user-defined times, and when hitting a surface; its complete trajectory; and the trajectory of its spin vector.

PENTrack is written in C++ and its object-oriented structure simplifies the implementation of new electromagnetic fields, particles, and physics processes. It is based on open-source libraries and freely available at \href{https://github.com/wschreyer/PENTrack.git}{\nolinkurl{github.com/wschreyer/PENTrack.git}}.

\section{Validation}

We performed two benchmark simulations to compare PENTrack with \textsc{Geant4} v10.2.2 and STARucn v1.2. In addition, we modeled two UCN experiments and were able to replicate their results using PENTrack.

\subsection{Comparison with other simulation tools}

\begin{figure}
\centering
\includegraphics[width=0.5\linewidth]{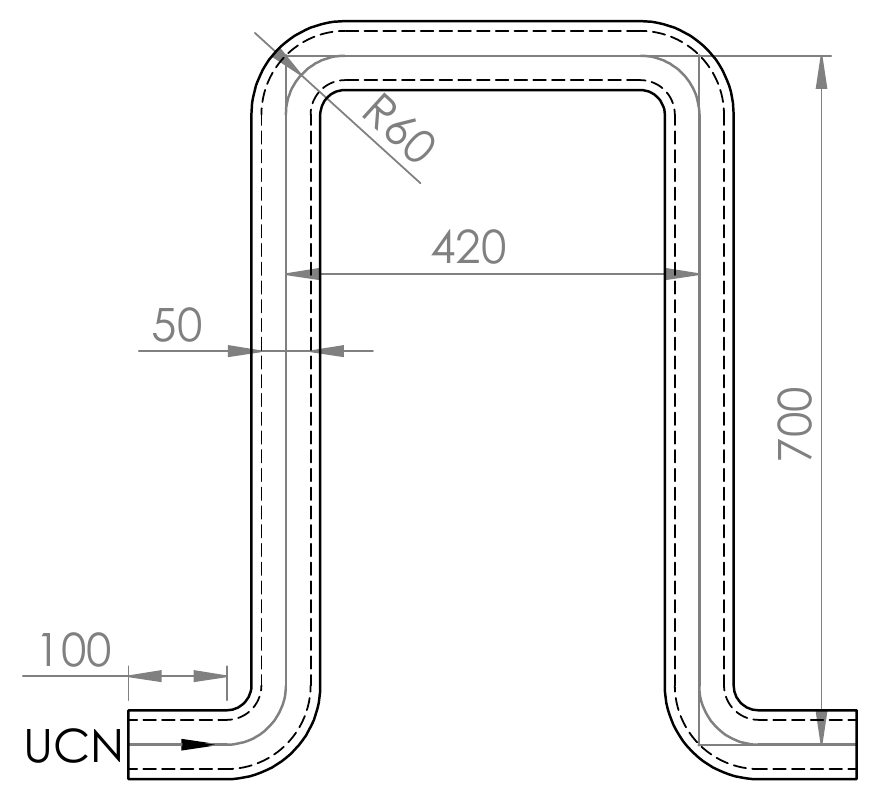}
\caption{Geometry of the benchmark simulation used to compare PENTrack, \textsc{Geant4}, and STARucn. Dimensions are given in millimeters.}
\label{fig:U}
\end{figure}

\begin{figure}
\includegraphics[width=\linewidth]{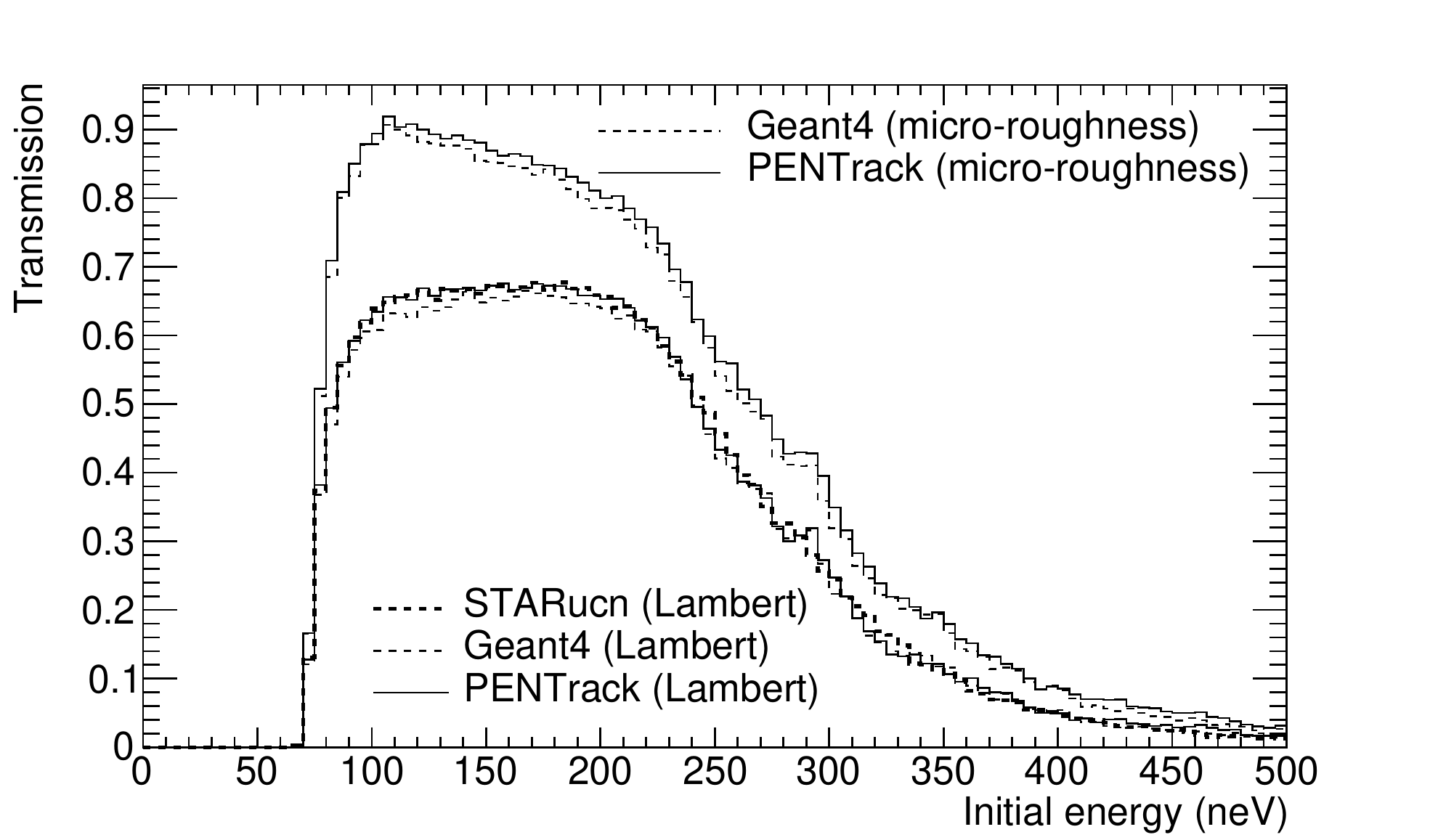}
\caption{Transmission of UCNs through a U-shaped guide (fig. \ref{fig:U}), simulated with PENTrack, \textsc{Geant4}, and STARucn. STARucn does not support microroughness reflection.}
\label{fig:STARucn_benchmark}
\end{figure}

\begin{figure}
	\centering
	\includegraphics[width=0.8\linewidth]{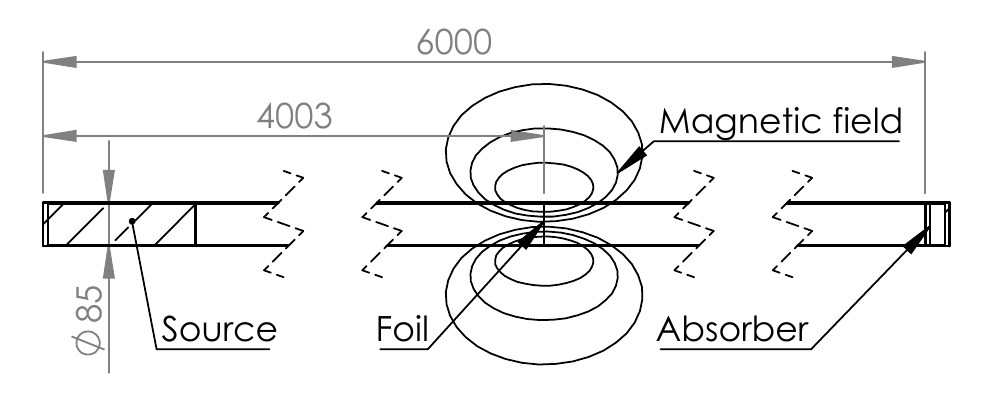}
	\caption{Geometry of the benchmark simulation used to compare PENTrack and \textsc{Geant4}. Dimensions are given in millimeters.}
	\label{fig:G4benchmark}
\end{figure}

\begin{figure}
	\centering
	\includegraphics[width=\linewidth]{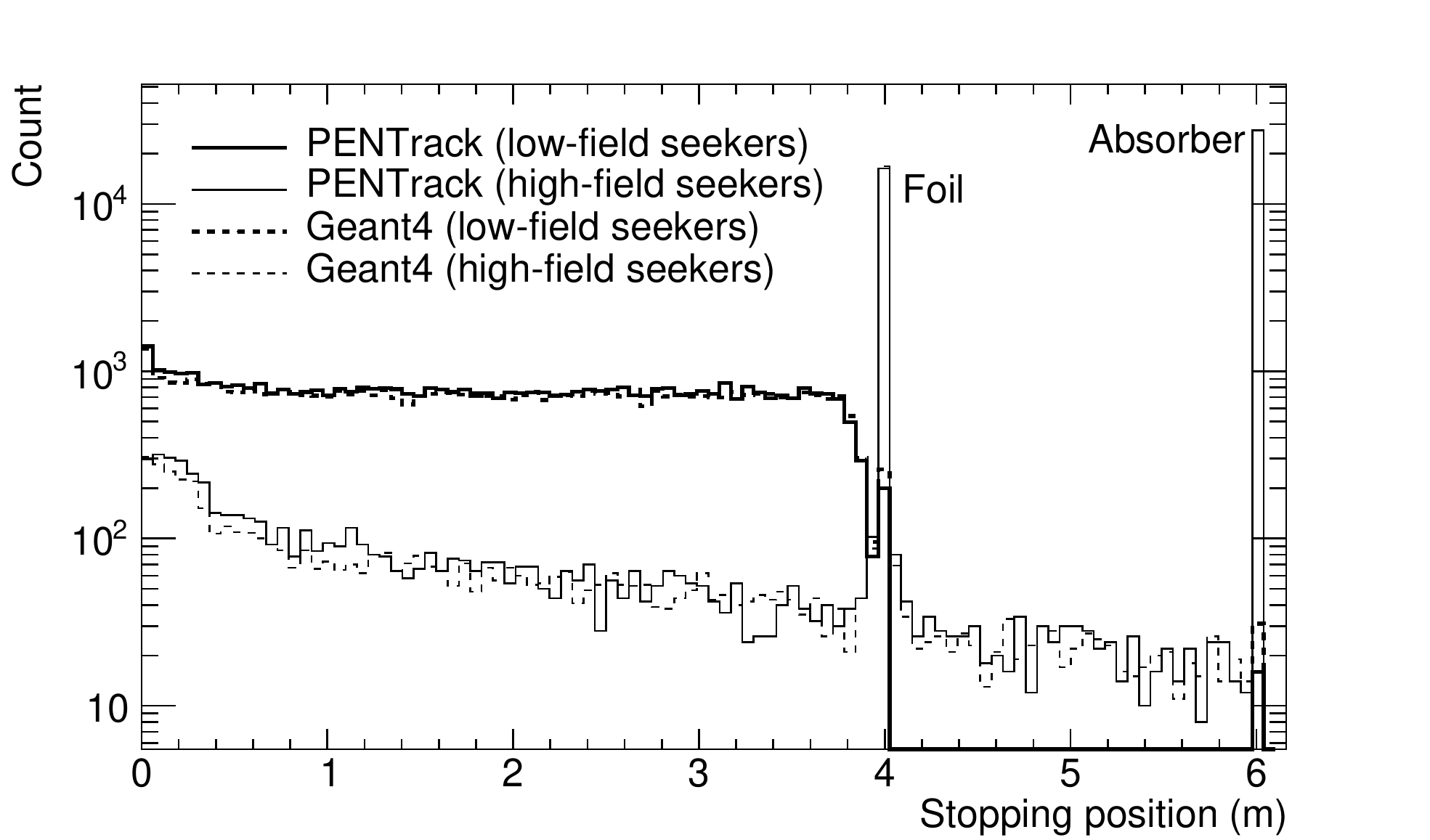}
	\caption{Positions where UCNs were absorbed by the guide, foil, or absorber seen in fig. \ref{fig:G4benchmark}, in simulations with PENTrack and \textsc{Geant4}.}
	\label{fig:geant4}
\end{figure}

The first benchmark simulates transport of UCN through a vertical, U-shaped UCN guide with a square cross section of \SI{5x5}{\centi\meter} shown in fig. \ref{fig:U}. We selected stainess steel with an optical potential of \SI{184 - 0.0184 i}{neV} as material of the guide and used either microroughness reflection\footnote{We assumed a roughness amplitude of \SI{2.6}{\nano\meter} and a correlation length of \SI{20}{\nano\meter} \cite{Atchison10}.} or a \num{5}-\% probability of diffuse Lambertian reflection. Both ends of the guide are covered with perfect absorbers. The simulations uncovered several flaws in the microroughness reflection implemented in \textsc{Geant4} v10.2.2, resulting in incorrect scattering distributions and probabilities. Once these flaws were corrected, the transmission of UCNs through the guide agreed very well among all three programs, with a slightly lower transmission in \textsc{Geant4} (fig. \ref{fig:STARucn_benchmark}).

The second benchmark simulates the trajectories of UCNs in a strong magnetic field and in matter. The geometry consists of a guide tube with a length of \SI{6}{\meter} and a diameter of \SI{85}{\milli\meter}, coated with diamond-like carbon (fig. \ref{fig:G4benchmark}). A cylindrically shaped source generates UCNs at one end. Four meters downstream of the source, a superconducting polarizer magnet generates an inhomogeneous magnetic field penetrating the guide tube. We placed an aluminium foil with a thickness of \SI{0.1}{\milli\meter} in the center of the field and a polyethylene absorber at the far end of the guide.

UCNs with one polarization state, so-called low-field seekers, are repelled by the strong magnetic field and cannot penetrate the magnetic barrier. They are mainly absorbed on the source side of the guide. UCNs with the other polarization state, called high-field seekers, are attracted to the strong magnetic field and accelerated towards the foil. They are mostly absorbed by the foil or the absorber at the far end.

The simulated results showed that \textsc{Geant4} did not correctly account for refraction when a UCN entered a material, resulting in too little absorption of UCNs in the foil. Once this flaw was corrected, the results of \textsc{Geant4} and PENTrack agreed very well (fig. \ref{fig:geant4}).

All corrections to \textsc{Geant4} will be included in version 10.3.

\subsection{Comparison with experiments}

\begin{figure}
\includegraphics[width=\linewidth]{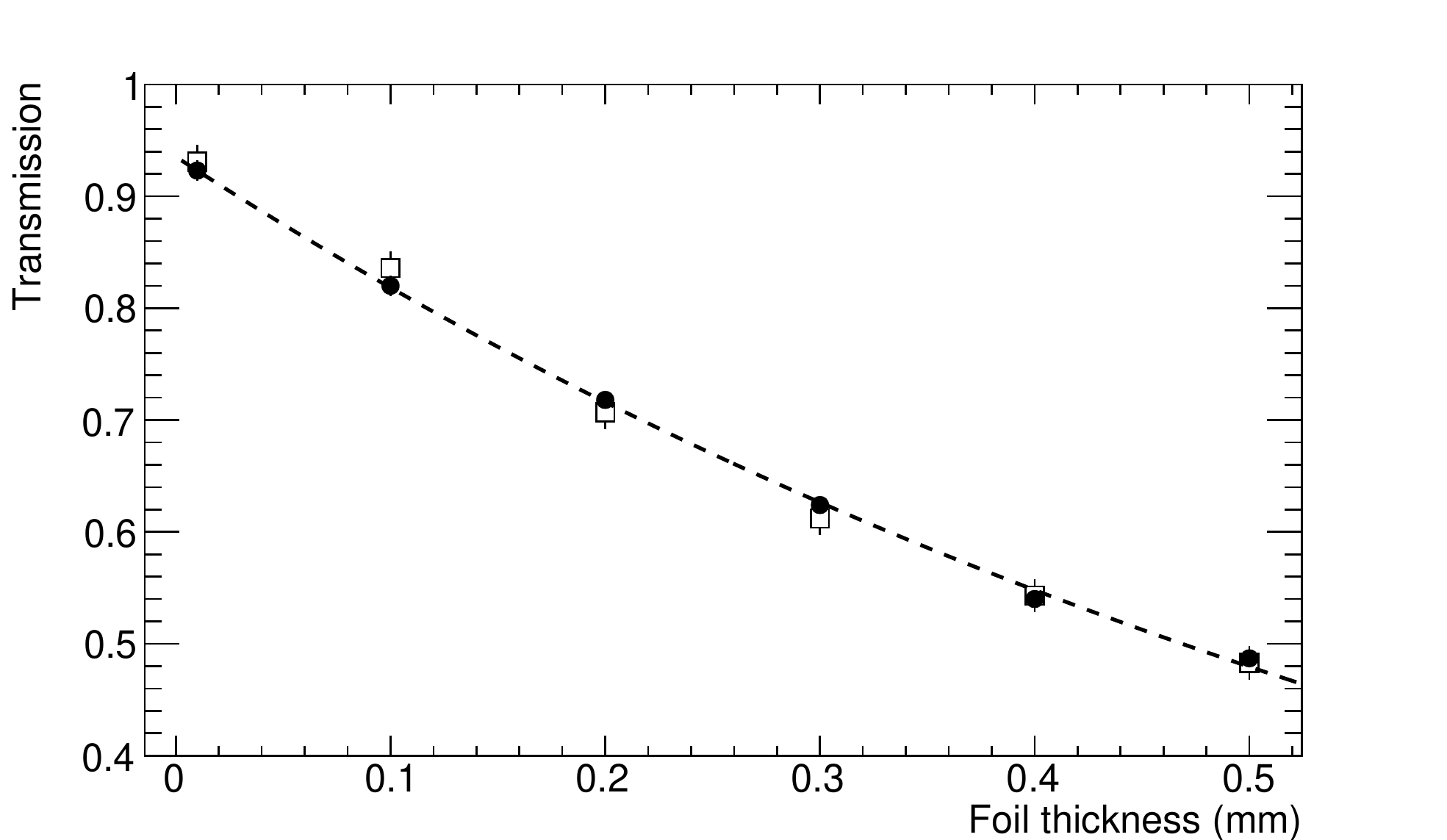}
\caption{UCN transmission through thin aluminium foils in experiment \cite{Atchison10Transmissionthroughmaterialfoils} (empty squares) and simulation (filled circles). The dashed line is an exponential fit to the simulated data.}
\label{fig:foiltrans}
\end{figure}

The first experiment we simulated measured transmission of UCN through thin foils of pure aluminium \cite{Atchison10Transmissionthroughmaterialfoils}. We imported their time-of-flight geometry and reproduced the quoted time-of-flight spectrum of UCN with an initially cosine-distributed angle between their velocity and the guide axis. We assigned a diffuse Lambert-scattering probability of \SI{10}{\percent} and the quoted loss cross section of \SI{229}{\barn} at \SI{6.2}{\meter\per\second} to the aluminium foils, resulting in an optical potential of \SI{54.1 - 0.00281 i}{\nano\electronvolt}. The simulated transmission rate of UCN through foils with different thicknesses agrees very well with the experimental data (fig. \ref{fig:foiltrans}). The mean free path of \SI{0.748 +- 0.016}{\milli\meter} determined from an exponential fit to the simulated data matches the experimental value of \SI{0.725 +- 0.009}{\milli\meter}.

\begin{figure}
\includegraphics[width=\linewidth]{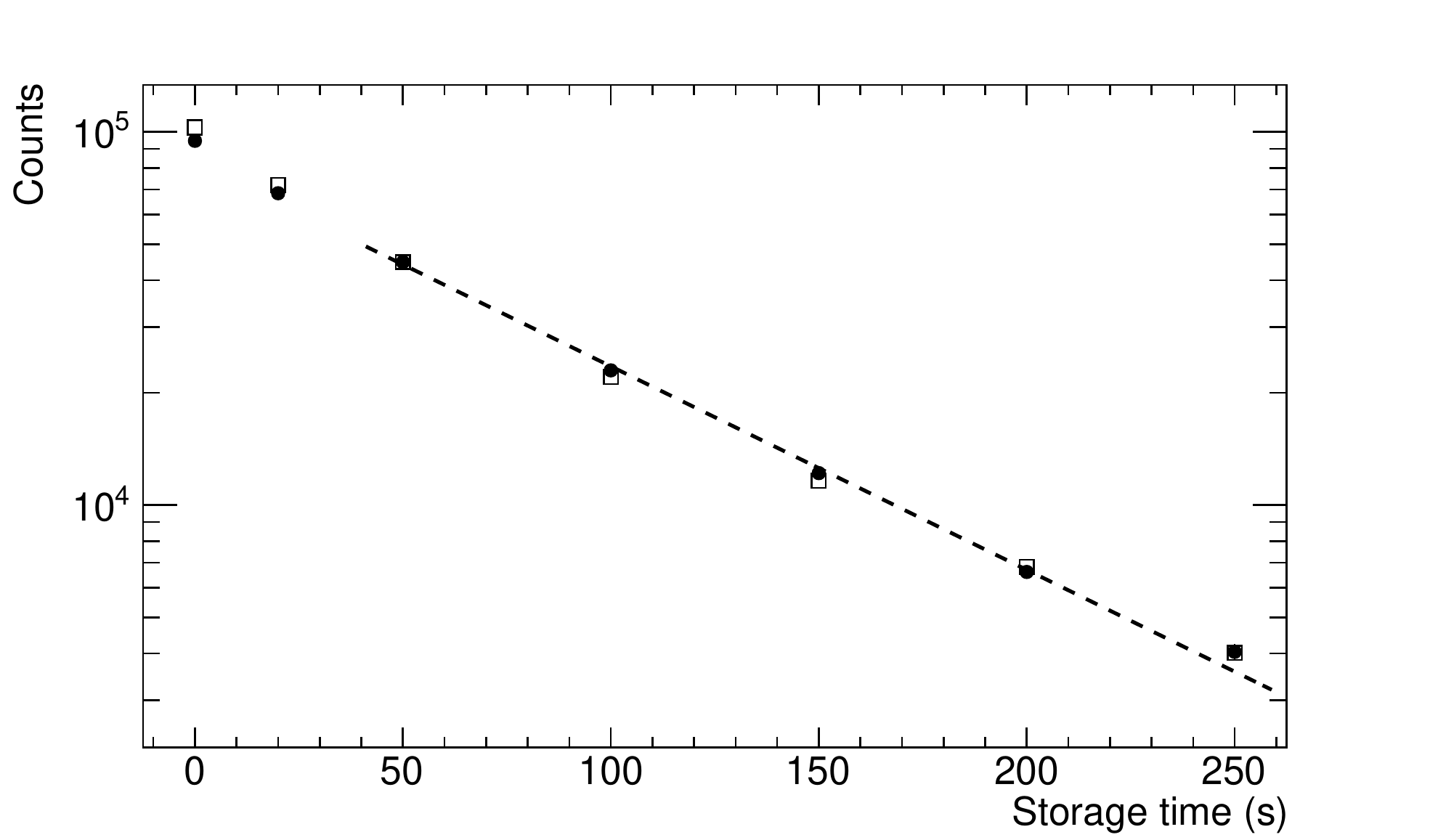}
\caption{Number of UCNs extracted from the UCN source at RCNP following different storage times. Experimental data from \cite{Masuda2012} is shown as empty squares, simulation results---scaled to the experimental data point following a storage time of \SI{50}{\second}---as filled circles. The dashed line is an exponential fit to the simulated data.}
\label{fig:storage_time}
\end{figure}

In a second simulation, we imported the geometry of the UCN source at the Research Center for Nuclear Physics (RCNP), Osaka University, as described in \cite{Masuda2012} and compared the resulting storage time of UCN with experimental data. The source is filled with superfluid helium at \SI{0.8}{\kelvin} and enclosed by walls coated with NiP. Since these walls are cold, we assumed that no inelastic scattering takes place and assigned an optical potential of \SI{213 - 0.0224 i}{\nano\electronvolt} to the walls. Following \cite{Golub1979}, we assumed a UCN-loss rate in the superfluid helium of \(v_\mathrm{n} n_\mathrm{He} \sigma_{l,\mathrm{He}} = \SI{0.00275}{\per\second}\), resulting in an optical potential of \((18.5 - 9.05 \cdot 10^{-10} \mathrm{i}) \, \si{\nano\electronvolt}\). Guides from the source to a UCN valve and a detector are made of stainless steel with an optical potential of \SI{183 - 0.0852 i}{\nano\electronvolt}. At the beginning of the simulation, the source generates UCNs with an energy spectrum proportional to \(\sqrt{E}\) between \num{0} and \SI{350}{\nano\electronvolt}. After a certain storage time, the valve opens and UCNs are extracted into the detector. The UCN lifetime in the source of \SI{79.6 +- 1.2}{\second}, determined from the measurements with storage times of \SI{50}{\second} or more, matches the experimental result of \SI{80.9 +- 0.4}{\second} very well (fig. \ref{fig:storage_time}).

\subsection{Comparison with analytical calculations}

\begin{figure}
\includegraphics[width=\linewidth]{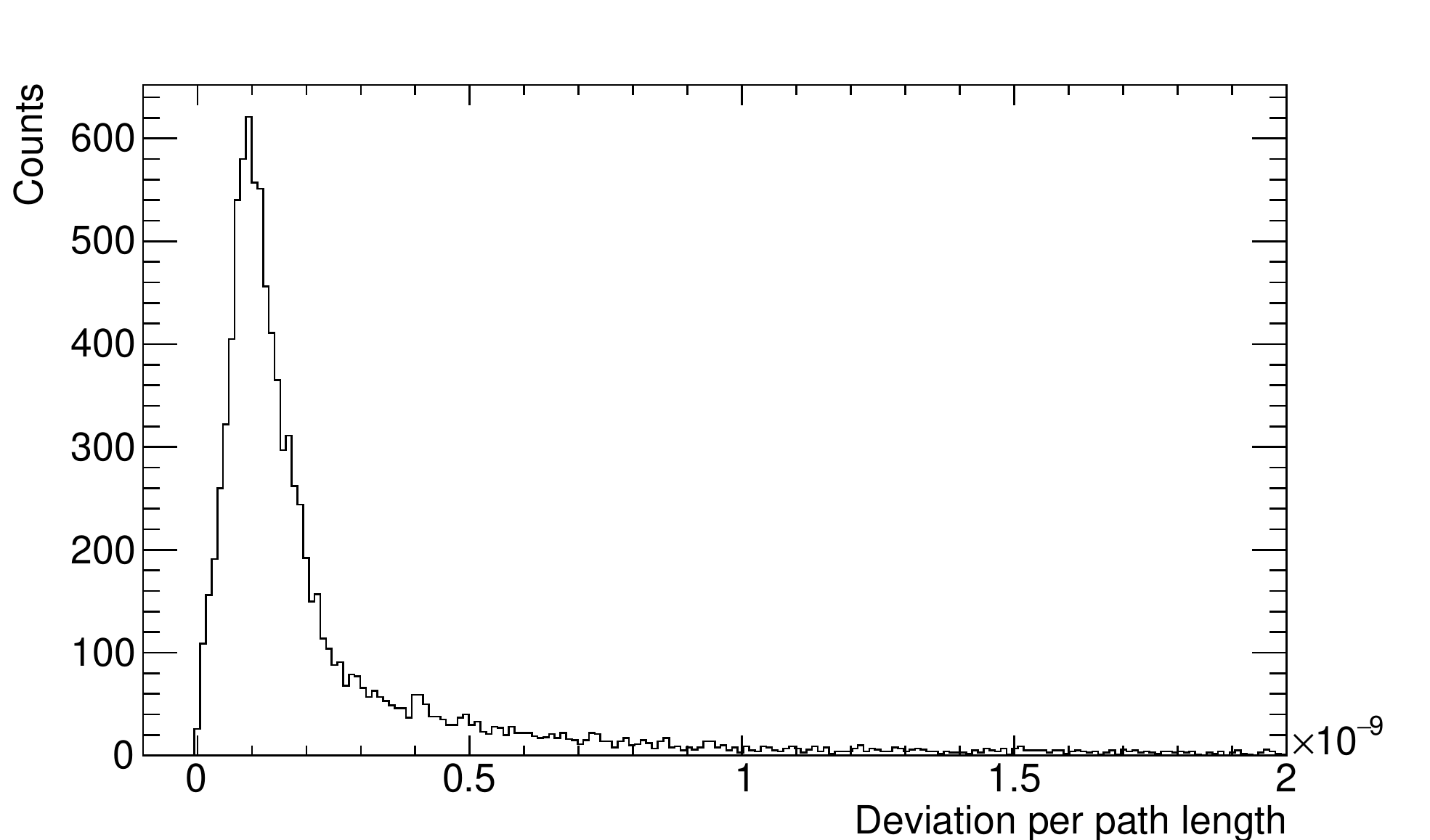}
\caption{Deviation of \num{10000} simulated electrons from their analytically calculated paths in orthogonal homogeneous electric and magnetic fields.}
\label{fig:electrondeviation}
\end{figure}

To validate the simulation of relativistic particles, we simulated beta-decay electrons with energies of up to \SI{780}{\kilo\electronvolt} and isotropic velocity distribution in orthogonal homogeneous electric and magnetic fields with strengths of \SI{10}{\kilo\volt\per\centi\meter} and \SI{0.1}{\tesla}. The resulting \(E \times B\) drift can be described analytically: In an inertial frame moving with velocity \(\mathbf{u} = \mathbf{E} \times \mathbf{B}/\mathbf{B}^2\) with respect to the fixed laboratory frame, the electrons follow a helical path along a reduced magnetic field \(\mathbf{B} \sqrt{1 - \mathbf{u}^2/c^2}\) \cite[ch. 12.3]{Jackson}. The difference between the analytical and simulated paths linearly increases with the path length and \SI{95}{\percent} of the simulated particles deviate by less than one nanometer after a flight path of one meter (fig. \ref{fig:electrondeviation}).

\begin{figure}
	\includegraphics[width=\linewidth]{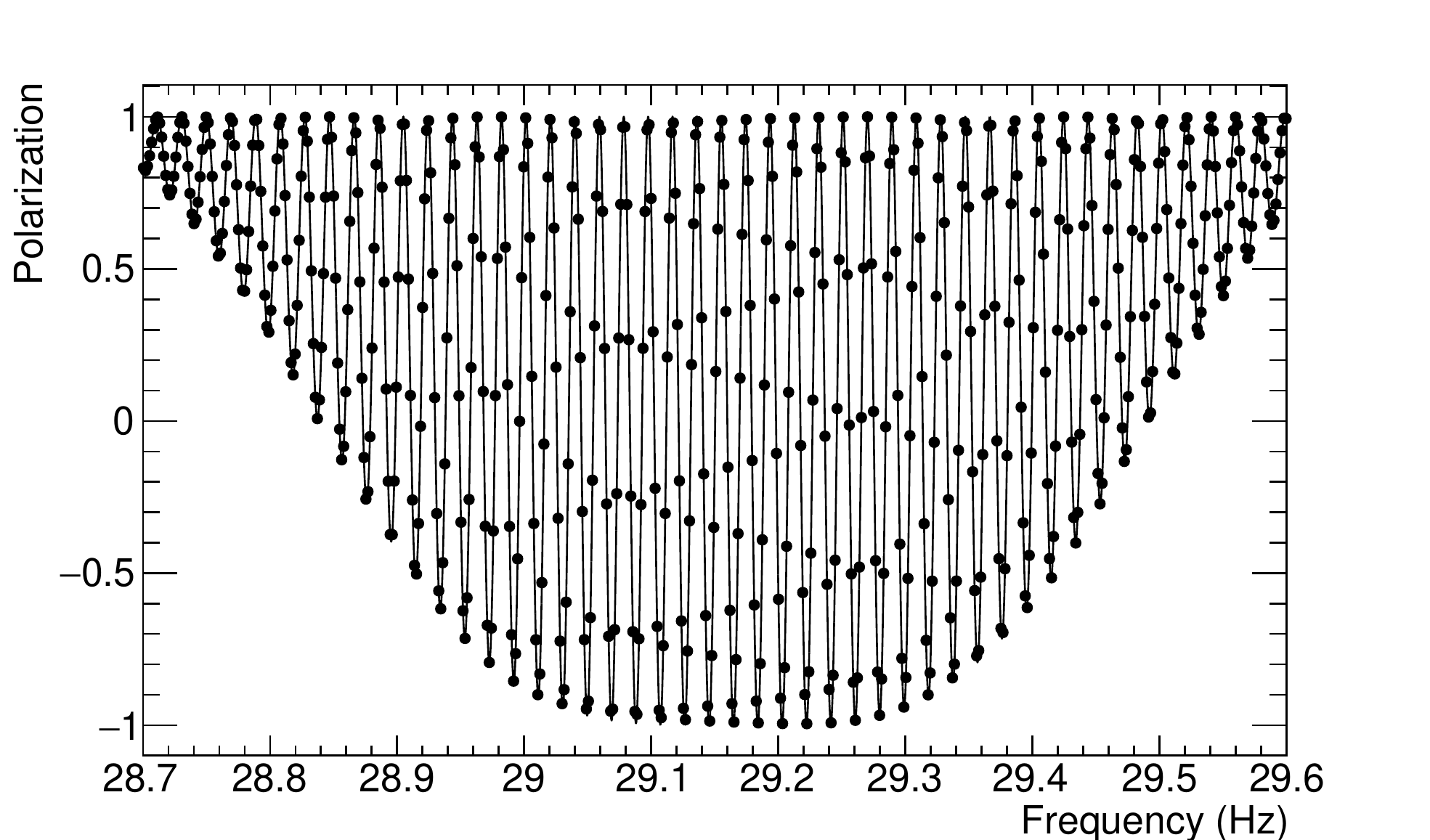}
	\caption{Polarization of a neutron spin after typical nEDM-experiment cycles with varying \(\pi/2\)-pulse frequency. The simulated results (dots) perfectly lie on the expected Ramsey pattern \cite{Gerthsen2015} (solid line).}
	\label{fig:Ramsey_fringes}
\end{figure}

To validate the simulation of spins, we simulated the spin of a neutron during typical nEDM-experiment cycles. nEDM measurements are based on Ramsey's method of separated oscillating fields \cite{Ramsey1950}. A short, oscillating magnetic-field pulse rotates the spins of polarized UCNs by an angle of \(\pi/2\). The spins are then left to precess freely in homogeneous magnetic and electric fields. After a certain time, another \(\pi/2\) pulse---in phase with the first pulse---again rotates the spin.

We simulated nEDM-experiment cycles using \(\pi/2\) pulses with an amplitude of \SI{10}{\nano\tesla} and varying frequency. After \SI{50}{\second} of free precession in a homogeneous 1-\si{\micro\tesla} magnetic field, the simulation generates the analytically calculated Ramsey pattern \cite{Gerthsen2015} (fig. \ref{fig:Ramsey_fringes}).

\section{Performance}

\begin{table}
\centering
\caption{Average number of UCNs simulated per second in the benchmark simulations using a single thread on an Intel Xeon E5520 processor.}
\begin{tabular}{lSSr}
\toprule
Simulation				& {PENTrack}& \textsc{Geant4}& {STARucn} \\
\midrule
U guide (Lambert)		& 530		& 56			& 2700	\\
U guide (MR)			& 170		& 51			& {--}	\\
tube, no field			& 16		& 3.1			& {--}	\\
tube, 3D field map		& 0.089		& 0.022			& {--}	\\
\bottomrule
\end{tabular}
\label{tab:perf}
\end{table}

Table \ref{tab:perf} summarizes the single-threaded processing speed of the three different tools during the benchmark simulations. For simple simulations of UCN transmission, STARucn offers the highest speed. In comparison, PENTrack is slower by a factor of five due to its flexible but computationally intensive geometry description. \textsc{Geant4} is three to ten times slower than PENTrack.

If material interactions are modeled with microroughness reflection, PENTrack is slowed down by a factor of three compared to simulations with purely Lambertian reflection. \textsc{Geant4} uses look-up tables for the microroughness distributions, which do not reduce processing speed but require an initialization time of \SI{460}{\second}. This time is not included in the calculations for table \ref{tab:perf}.

To speed up simulations of large numbers of particles, PENTrack was specifically designed to be run in several parallel instances on multi-core processors and computing clusters. One can assign a job number to each instance via command line, which is prepended to the corresponding output files. All output files can be merged with dedicated tools provided with PENTrack.

\section{Example applications}

\subsection{PENeLOPE}

\begin{figure}
\includegraphics[width=\linewidth]{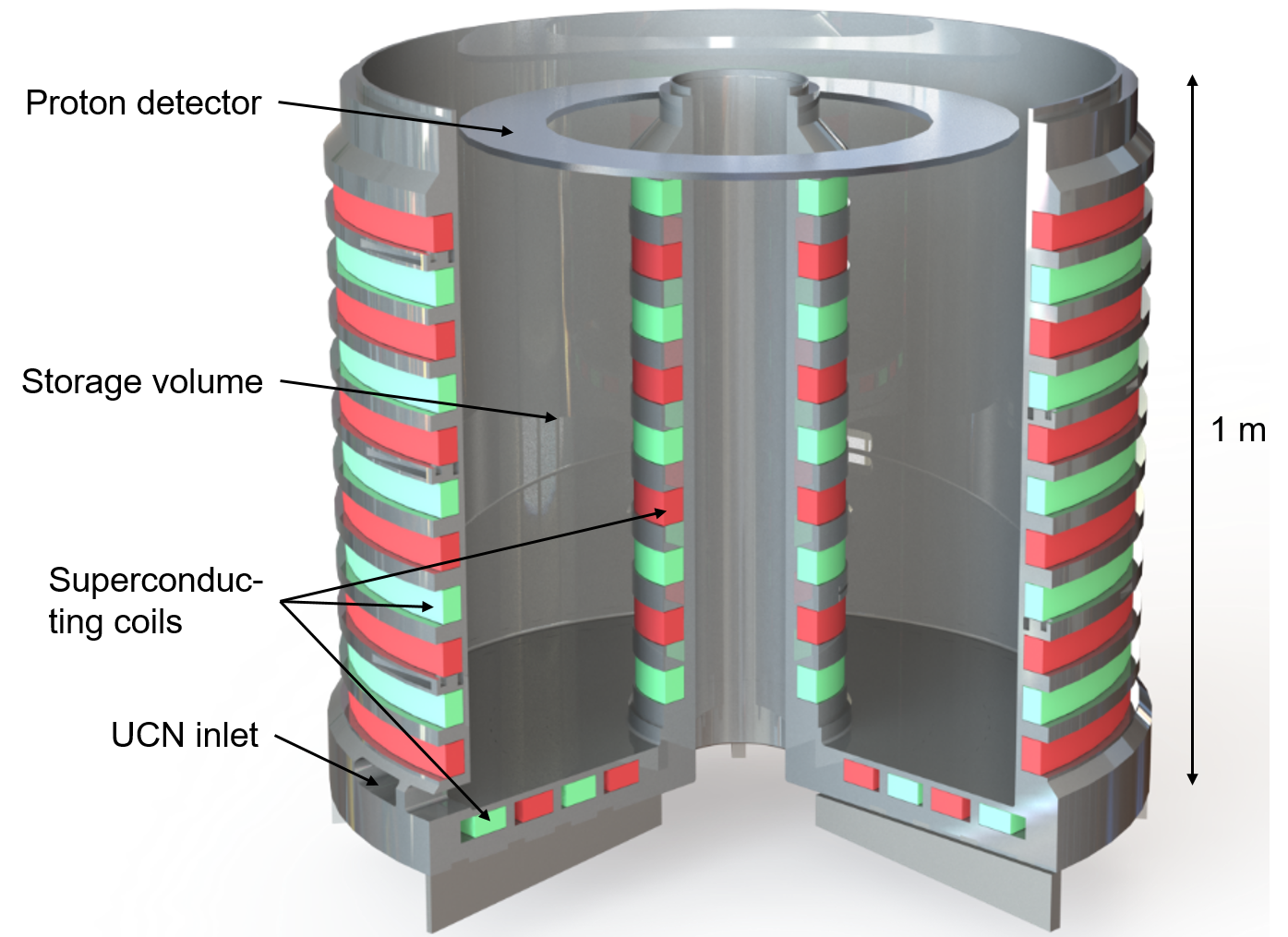}
\caption{Rendering of PENeLOPE's storage volume.}
\label{fig:PENeLOPE}
\end{figure}

Using PENTrack, we were able to simulate a complete measurement cycle of the PENeLOPE experiment. PENeLOPE uses a large, superconducting magnet (fig. \ref{fig:PENeLOPE}) to trap UCNs and measure the neutron lifetime.

At the beginning of each measurement cycle, UCNs are filled into the storage volume for \SI{200}{\second}, achieving \SI{95}{\percent} of saturation. A valve then closes the storage volume and UCNs are trapped by its walls made from stainless steel. For another \SI{200}{\second}, absorbers in the storage volume remove UCNs with energies high enough to overcome the minimum trapping potential during magnetic storage of \SI{115}{\nano\electronvolt}. After this cleaning stage, the superconducting magnet is ramped up within \SI{100}{\second} and low-field seekers become trapped by the magneto-gravitational potential
\begin{equation}
U_\mathrm{m} = \mu_\mathrm{n} \left| \mathbf{B} \right| + m_\mathrm{n} \left| \mathbf{g} \right| z.
\end{equation}
During magnetic storage, a proton detector at the top of the storage volume can directly observe the decay rate of the trapped UCNs, from which we can determine their lifetime in the trap. After a predefined storage time, the magnet is ramped down again and the remaining UCNs are counted by a UCN detector, providing a second measurement of their lifetime in the trap.

In our simulations, ultracold neutrons were created at the converter surface of a UCN source and transported through UCN guides to the experiment. The results allowed us to optimize the vertical position of the experiment, the guide geometry, and the filling time with respect to the number of stored UCNs.

\begin{figure}
	\includegraphics[width=\linewidth]{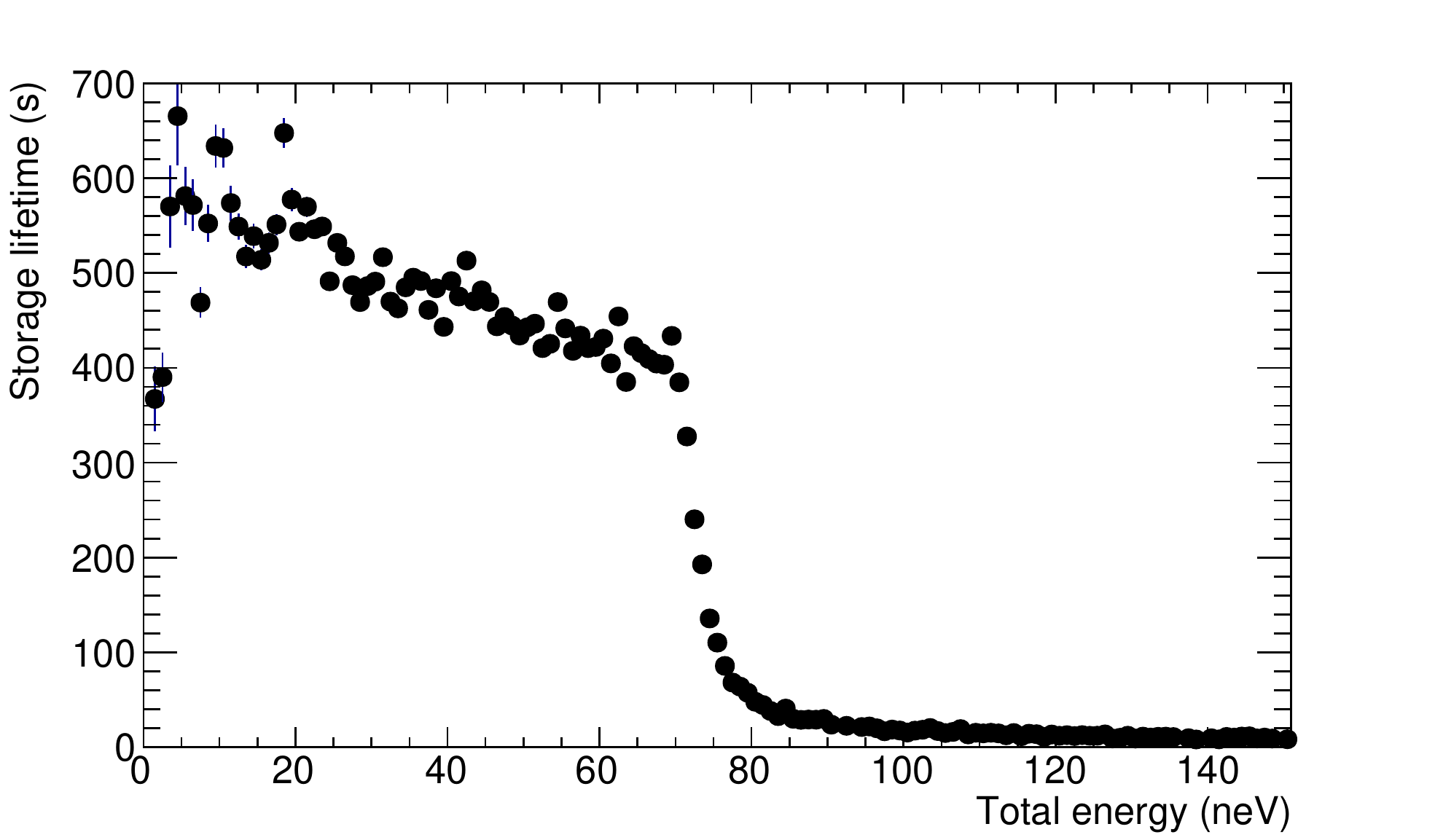}
	\caption{Storage lifetime of UCNs in PENeLOPE's storage volume without magnetic field. A polyethylene absorber at a height of \SI{0.7}{\meter} above the bottom of the storage volume reduces the storage lifetime of UCNs with total energies above \SI{70}{\nano\electronvolt}.}
	\label{fig:cleaning}
\end{figure}

Simulations of the cleaning stage allowed us to optimize the geometry of the UCN inlet to shorten the time required to remove UCNs with energies high enough to overcome the magneto-gravitational trapping potential (fig. \ref{fig:cleaning}).

\begin{figure}
	\includegraphics[width=\linewidth]{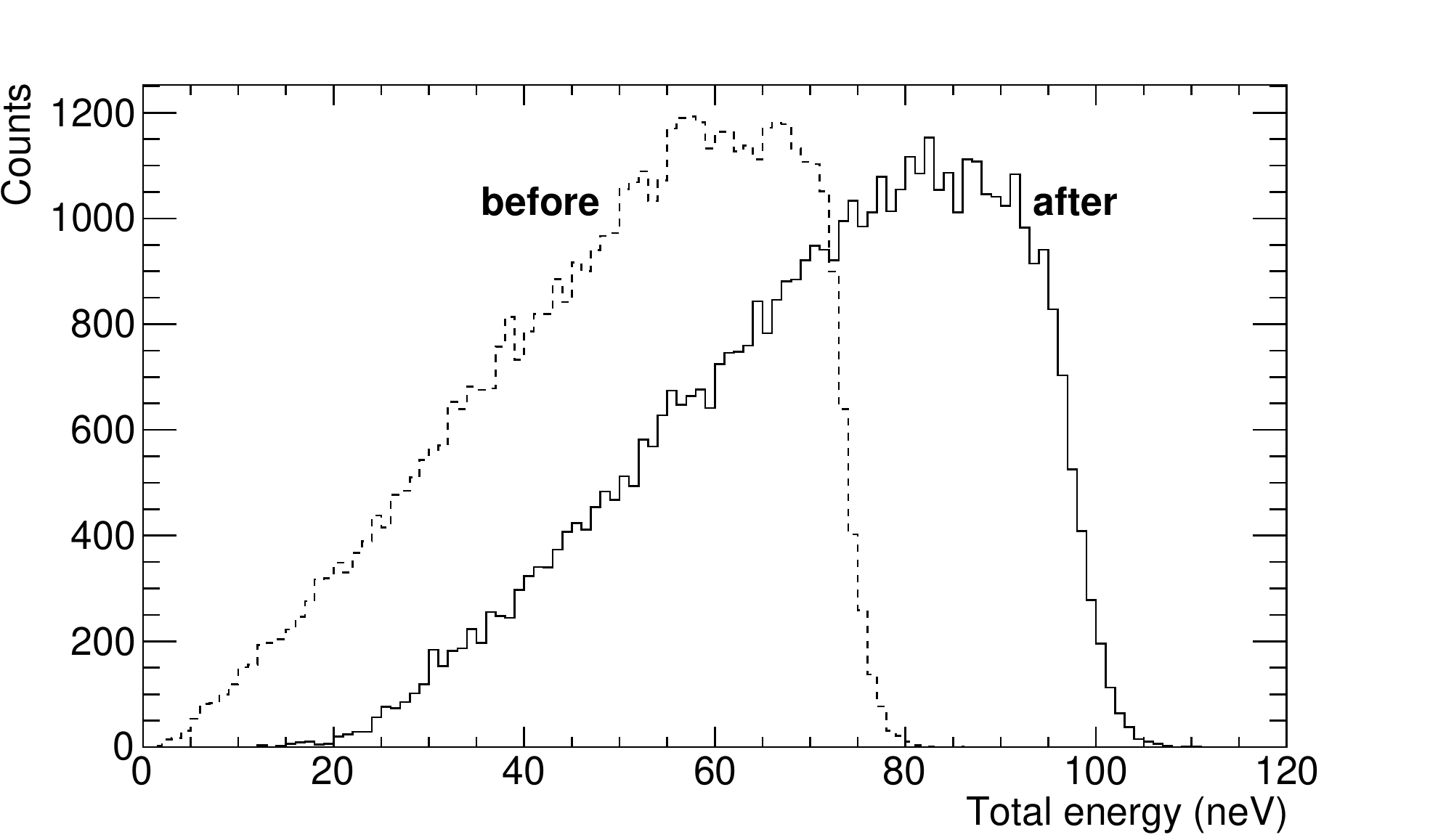}
	\caption{Total energy of low-field seekers in PENeLOPE's storage volume, before (dashed line) and after (solid line) ramping up the magnet.}
	\label{fig:heating}
\end{figure}

\begin{figure}
	\centering
	\includegraphics[width=0.6\linewidth]{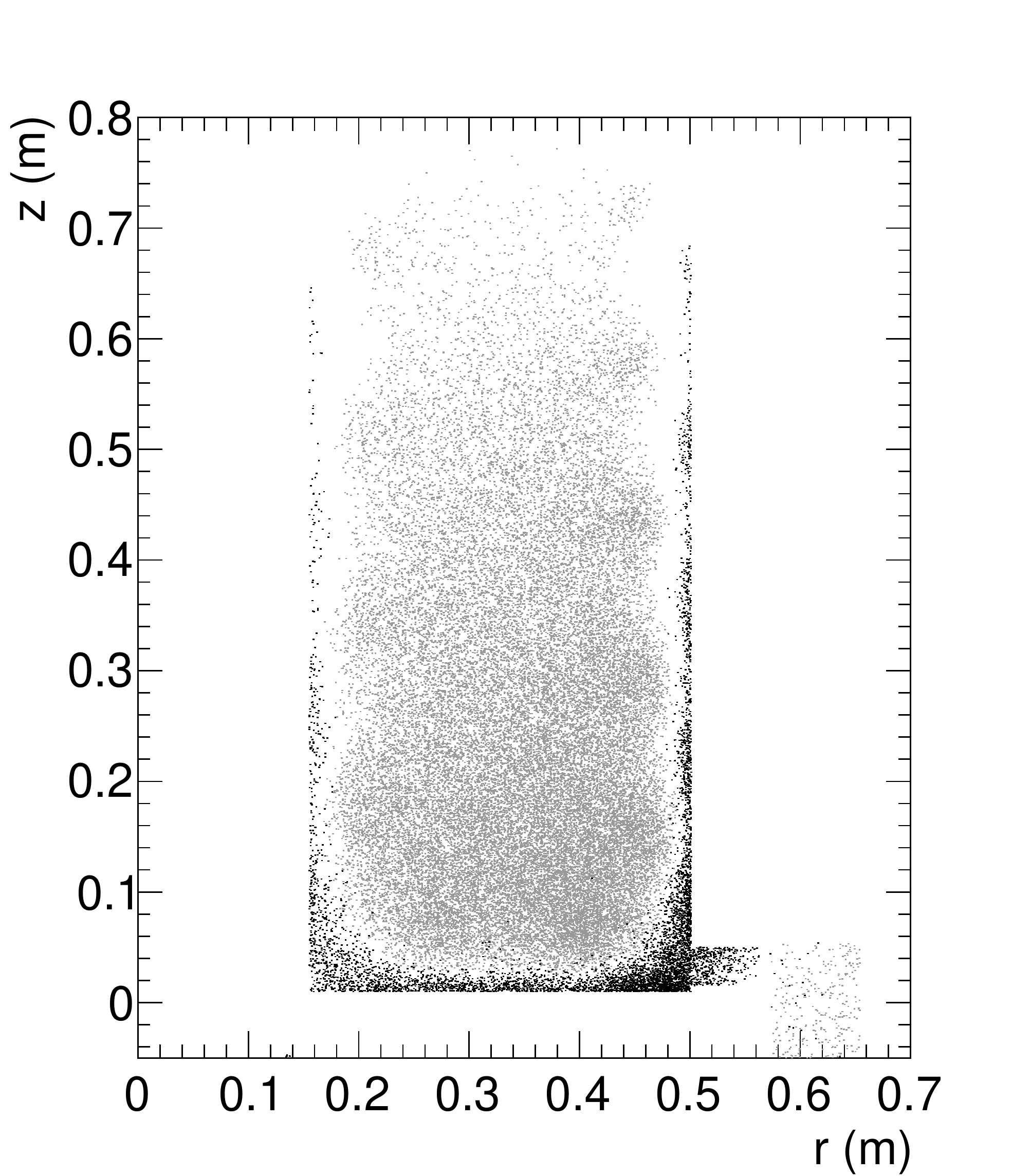}
	\caption{Cylindrically projected distribution of low-field seekers (gray) and high-field seekers (black) in PENeLOPE's storage volume after ramping up the magnet.}
	\label{fig:neutdist}
\end{figure}

While the superconducting magnet is ramped up, the slowly increasing magnetic potential increases the total energy of low-field seekers, \(E + U_\mathrm{m}\), which could push their total energy above the trapping potential (fig. \ref{fig:heating}). At the same time, high-field seekers are accelerated towards the walls and undergo many wall collisions (fig. \ref{fig:neutdist}). Both effects would introduce losses of UCNs and their lifetime in the trap would become shorter than the inherent neutron lifetime. From our simulations, we estimated the energy increase of low-field seekers during ramping. We also determined the loss rate of high-field seekers during magnetic storage and tested strategies to remove them from the storage volume.

\begin{figure}
	\includegraphics[width=\linewidth]{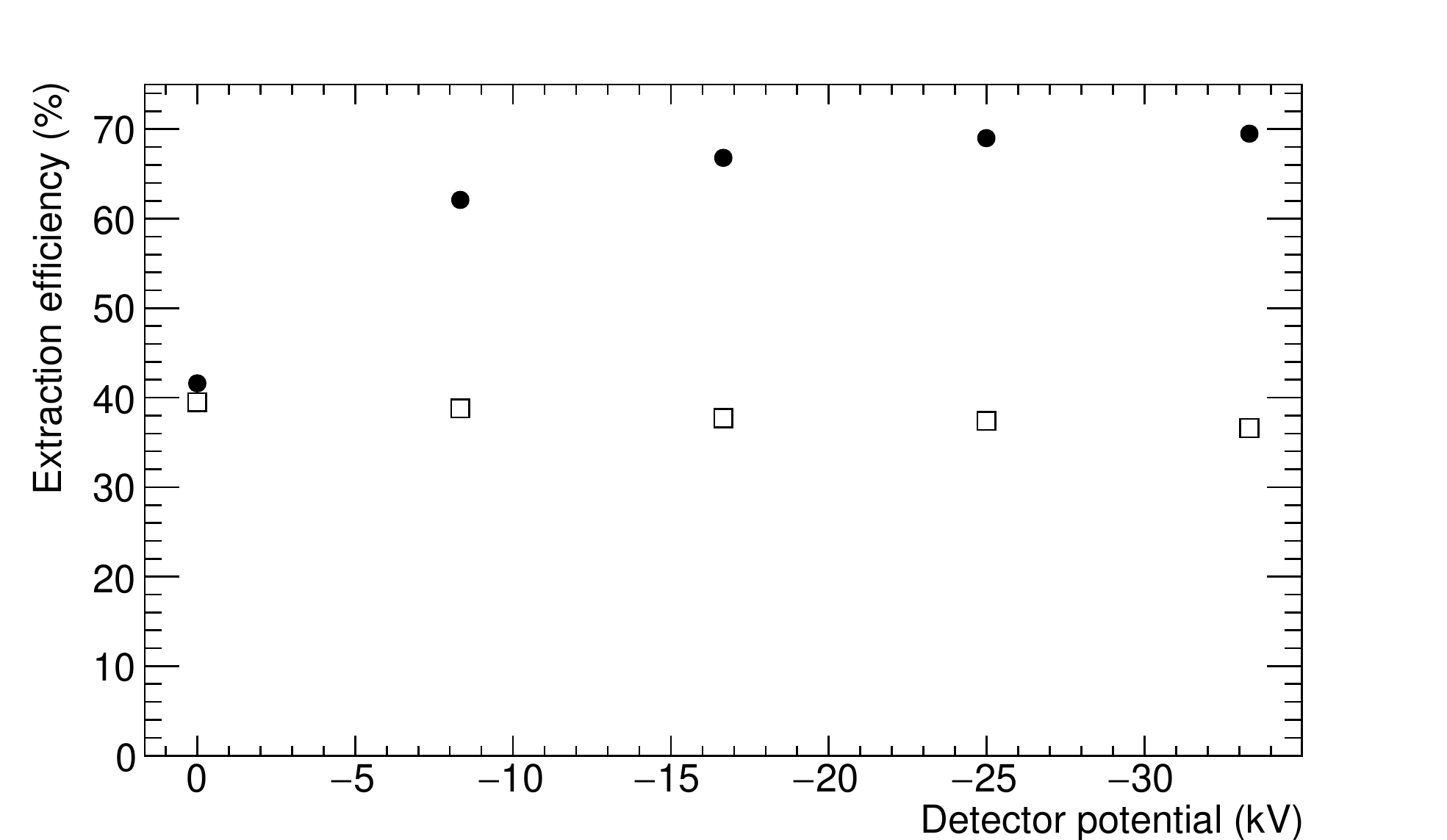}
	\caption{Fraction of protons (filled circles) and electrons (empty squares) from decays of magnetically stored low-field seekers reaching the detector.}
	\label{fig:extreff}
\end{figure}

Simulations of the trajectories of protons and electrons from decays of trapped low-field seekers during magnetic storage showed that a voltage of at least \SI{-25}{\kilo\volt} should be applied to the proton detector to efficiently extract and detect the decay protons with energies below \SI{0.75}{\kilo\electronvolt} (fig. \ref{fig:extreff}). The decay electrons can hardly be influenced due to their much higher energy of up to \SI{782}{\kilo\electronvolt}.

\subsection{Geometric phases in nEDM experiments}

\begin{figure}
	\includegraphics[width=\linewidth]{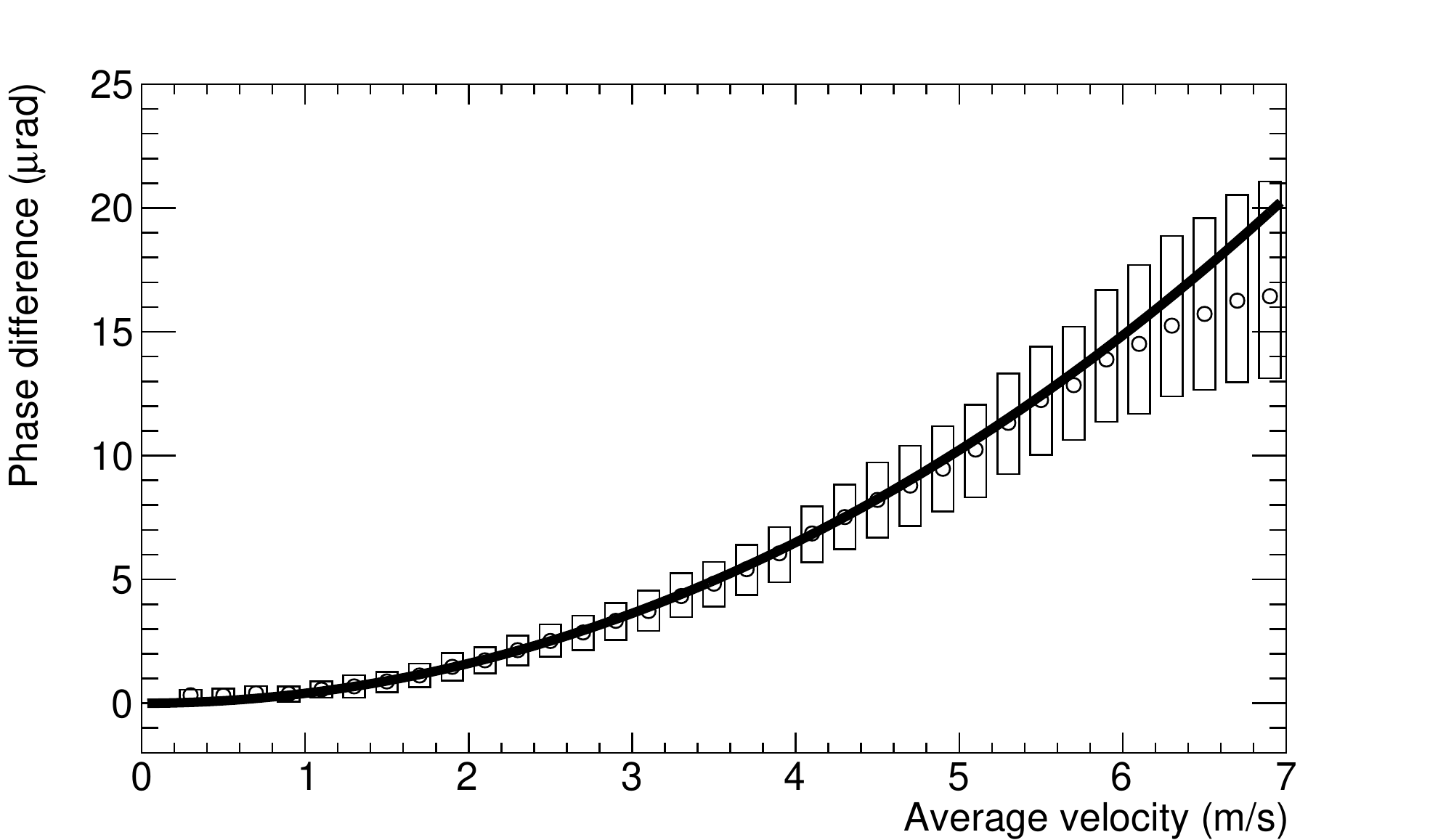}
	\caption{Distribution of phase differences between spins of pairs of UCNs with identical trajectories in opposite electric fields. The average of the simulated differences (circles) can be calculated analytically for low velocities (solid line) \cite[eq. (29)]{Pendlebury2004}. The boxes indicate the spread of the distribution from the first to the third quartile.}
	\label{fig:geomphase}
\end{figure}

In nEDM experiments, due to the electric field, a hypothetical electric dipole moment would slightly shift the precession frequency of the stored neutrons' spins. If the electric field is inverted, this shift is also inverted and a small phase difference between spins in opposite electric fields is accumulated over the free-precession time. A major uncertainty in nEDM experiments is caused by geometric phases, which can mimic the effect of an electric dipole moment. As shown in \cite{Pendlebury2004}, such geometric phases can arise due to small gradients in the magnetic field combined with the relativistic \(\dot{\mathbf{x}} \times \mathbf{E}\) term in the BMT equation (\ref{eq:vcrossE}). To make this tiny effect visible, we simulated pairs of UCNs with identical trajectories subjected to electric fields with opposite direction. The UCNs were stored in a cylindrical nEDM chamber with a radius of \SI{20}{\centi\meter} and a height of \SI{10}{\centi\meter}. It was placed in a magnetic field with a strength of \SI{1}{\micro\tesla} and a rotationally symmetric vertical gradient of \SI{10}{\pico\tesla\per\centi\meter}, and an electric field of \SI{+-10}{\kilo\volt\per\centi\meter}. Since the magnetic field has to obey the Maxwell equations, a vertical gradient leads to additional radial field components \cite{Pendlebury2004}. As shown in fig. \ref{fig:geomphase}, the spins of the stored UCNs accumulate a net phase difference, mimicking an electric dipole moment depending on the average velocity---although the UCNs have a random spatial distribution, have an isotropic velocity distribution, and undergo diffuse reflection on the chamber walls. This nicely replicates the calculations and simulations performed by \cite{Pendlebury2004}.

\section{Conclusions}

PENTrack is a tool that allows comprehensive simulations of neutron-lifetime and nEDM experiments---including UCN transport, UCN storage in material bottles and magnetic traps, spin precession of neutrons and co-magnetometer atoms, and tracking of protons and electrons in electromagnetic fields. It provides a flexible configuration interface and allows to load complex electromagnetic fields and geometries from FEM and CAD software.

Detailed comparisons of results obtained with PENTrack, STARucn v1.2, and \textsc{Geant4} v10.2.2 showed very good agreement with STARucn and uncovered several flaws in \textsc{Geant4}. STARucn offers higher speeds than PENTrack, but lacks support for microroughness reflection and magnetic fields. The speed of PENTrack is limited by its flexible geometry import, which, however, makes PENTrack much more suitable for implementing complicated experiment geometries. The very general particle-simulation framework \textsc{Geant4} offers similar functionality but has limited performance.

\section{Acknowledgements}

This work was supported by priority program SPP1491 ``Precision experiments in particle- and astrophysics with cold and ultracold neutrons'' of Deutsche Forschungsgemeinschaft and the Cluster of Excellence Exc153 ``Origin and Structure of the Universe''.

\bibliographystyle{elsarticle-num}
\bibliography{bibexport}

\end{document}